\documentclass[10pt,conference,compsocconf,letterpaper]{./sty/IEEEtran}
\makeatletter
\def\ps@headings{%
\def\@oddhead{\mbox{}\scriptsize\rightmark \hfil \thepage}%
\def\@evenhead{\scriptsize\thepage \hfil \leftmark\mbox{}}%
\def\@oddfoot{}%
\def\@evenfoot{}}
\makeatother
\usepackage{amsmath,theorem}
\usepackage[british]{babel}
\usepackage{graphics,color}
\usepackage{cite}
\usepackage{amssymb}
\usepackage[ruled,vlined,titlenotnumbered]{algorithm2e}
\usepackage{stfloats}
\usepackage[caption=false,font=footnotesize,subrefformat=parens]{subfig}
\usepackage{color}
\usepackage{balance}

\ifCLASSINFOpdf
   \usepackage[pdftex]{graphicx}
   \graphicspath{{./pdf/}{./jpeg/}{./fig/}}
   \DeclareGraphicsExtensions{.pdf,.jpeg,.png}
\else
   \usepackage[dvips]{graphicx}
   \graphicspath{{./eps/}}
   \DeclareGraphicsExtensions{.eps}
\fi

\def\fullfigwidth{1.0\textwidth}
\def\1figwidth{0.9\textwidth}
\def\2figwidth{0.45\textwidth}
\def\3figwidth{0.3\textwidth}
\def\4figwidth{0.23\textwidth}

\DeclareMathOperator*{\argmin}{arg\,min}

\usepackage{colortbl}

\newcommand{\term}[1]{{\it #1}}
\newcommand{\empha}[1]{{\bf #1}}

\newcommand{\bm}[1]{\boldsymbol #1}
\newcommand{\ef}[1]{#1 \!\!\! #1}

\def\sysname{DorFin}

\newtheorem{obsv}{Observation}

\ifodd 0
\newcommand{\rev}[1]{{\color{blue}#1}} 
\newcommand{\com}[1]{\textbf{\color{red}(COMMENT: #1)}} 
\else
\newcommand{\rev}[1]{#1}
\newcommand{\com}[1]{}
\fi

\begin{document}
\title{
DorFin: WiFi Fingerprint-based Localization Revisited
}
\author{\IEEEauthorblockN{Chenshu Wu\IEEEauthorrefmark{1},
Zheng Yang\IEEEauthorrefmark{1},
Zimu Zhou\IEEEauthorrefmark{2},
Yunhao Liu\IEEEauthorrefmark{1}
and Mingyan Liu\IEEEauthorrefmark{3}
} 
\IEEEauthorblockA{\IEEEauthorrefmark{1}School of Software and TNList, Tsinghua University}
\IEEEauthorblockA{\IEEEauthorrefmark{2}CSE, Hong Kong University of Science \& Technology}
\IEEEauthorblockA{\IEEEauthorrefmark{3}Department of EECS, University of Michigan}
\IEEEauthorblockA{{\{wu, yang, zimuzhou, yunhao\}@greenorbs.com}, {mingyan@eecs.umich.edu}}
}

\maketitle

\begin{abstract} 
Although WiFi fingerprint-based indoor localization is attractive, its accuracy remains a primary challenge especially in mobile environments. Existing approaches either appeal to physical layer information or rely on extra wireless signals for high accuracy. 
In this paper, we revisit the RSS fingerprint-based localization scheme and reveal crucial observations that act as the root causes of localization errors, yet are surprisingly overlooked or even unseen in previous works. Specifically, we recognize APs' diverse discrimination for fingerprinting a specific location, observe the RSS inconsistency caused by signal fluctuations and human body blockages, and uncover the RSS outdated problem on commodity smartphones. 
Inspired by these insights, we devise a discrimination factor to quantify different APs' discrimination, incorporate robust regression to tolerate outlier measurements, and reassemble different fingerprints to cope with outdated RSSs. 
Combining these techniques in a unified solution, we propose \sysname, a novel scheme of fingerprint generation, representation, and matching, which yields remarkable accuracy without incurring extra cost. 
Extensive experiments demonstrate that \sysname~achieves mean error of 2 meters and more importantly, bounds the 95th percentile error under 5.5 meters; these are about 56\% and 69\% lower, respectively, compared with the state-of-the-art schemes such as Horus and RADAR.


\end{abstract}

\section{Introduction}

\begin{figure*}[t]
	\centering
	\begin{minipage}[t]{\3figwidth}\centering
		\includegraphics[width=\fullfigwidth]{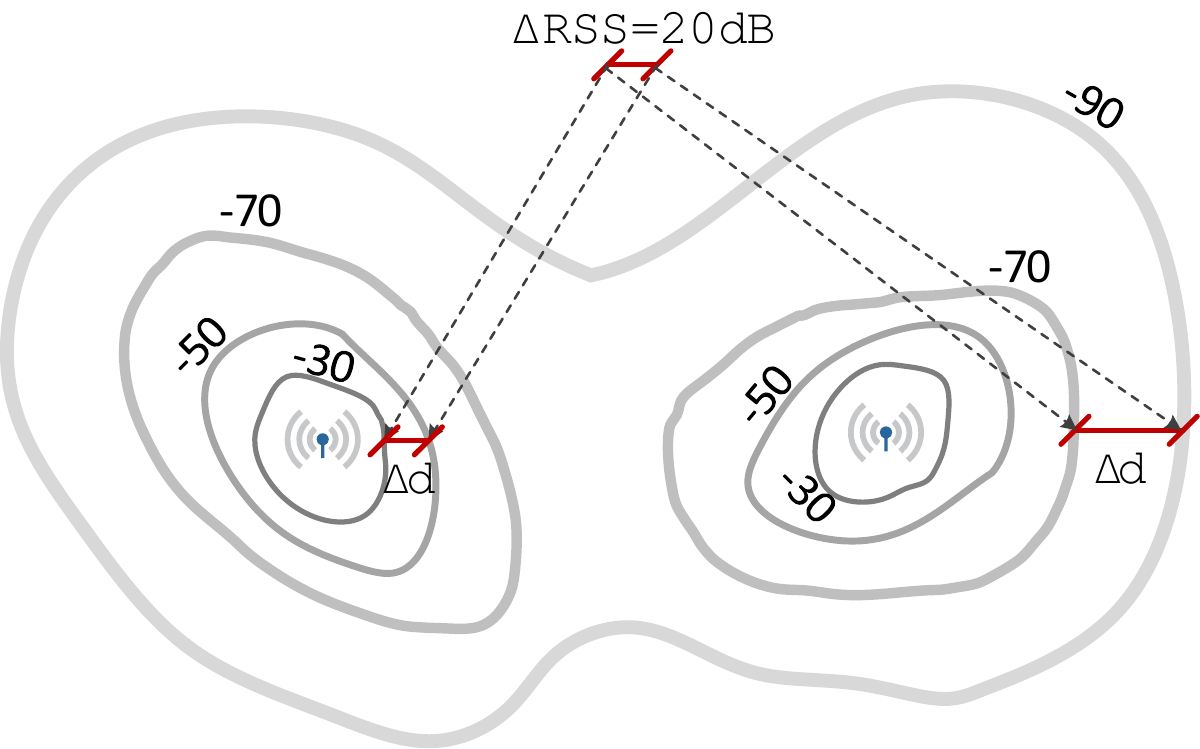}
  		\caption{Discrimination Diversity}
  		\label{fig:discrimination}
	\end{minipage}
	\hspace{0.2in}
	\begin{minipage}[t]{\3figwidth}\centering
		\centering
  		\includegraphics[width=\fullfigwidth]{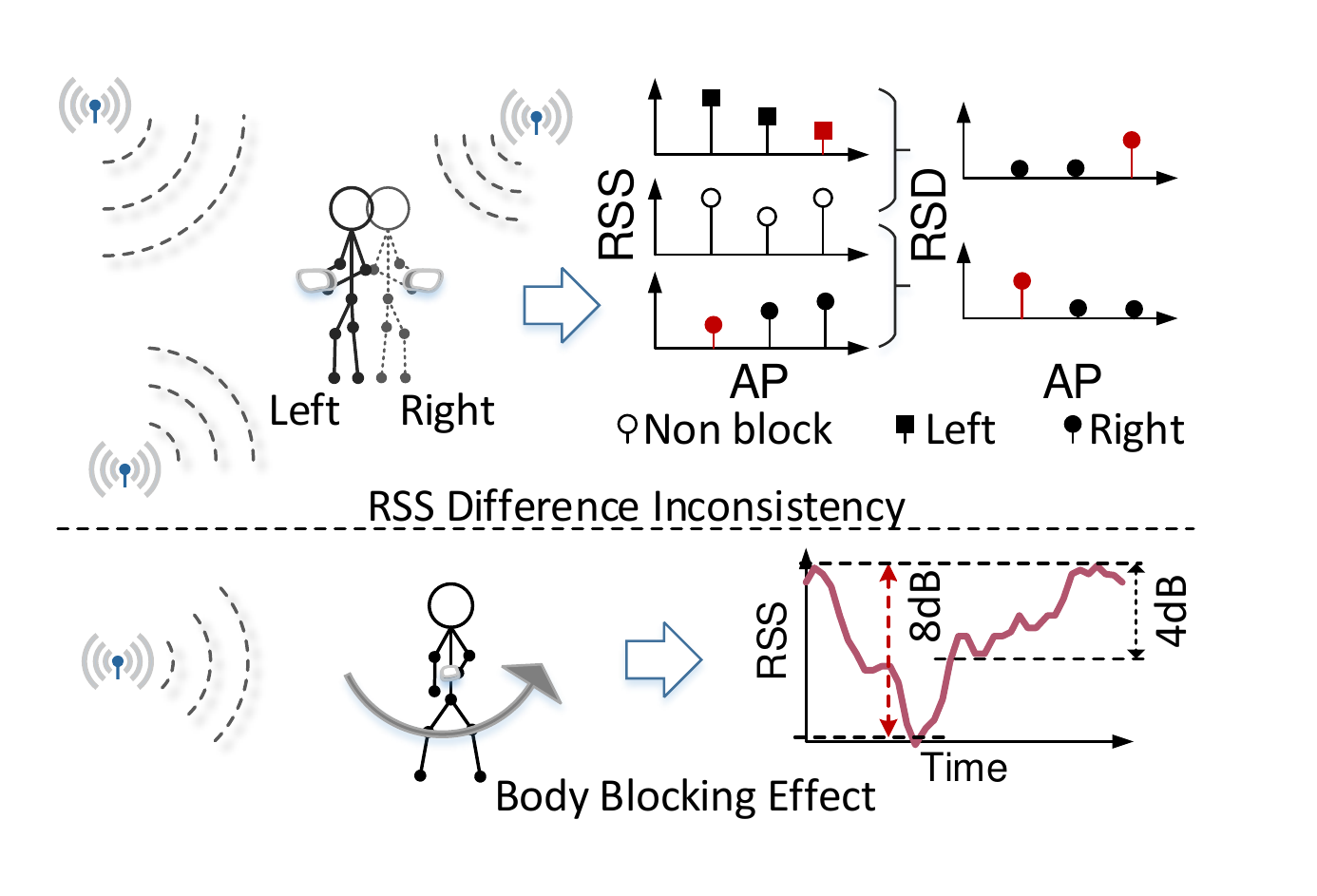}
  		\caption{Fingerprint Inconsistency}
 		\label{fig:bodyblock}
	\end{minipage}	
	\hspace{0.2in}
	\begin{minipage}[t]{\3figwidth}\centering
		\centering
  		\includegraphics[width=\fullfigwidth]{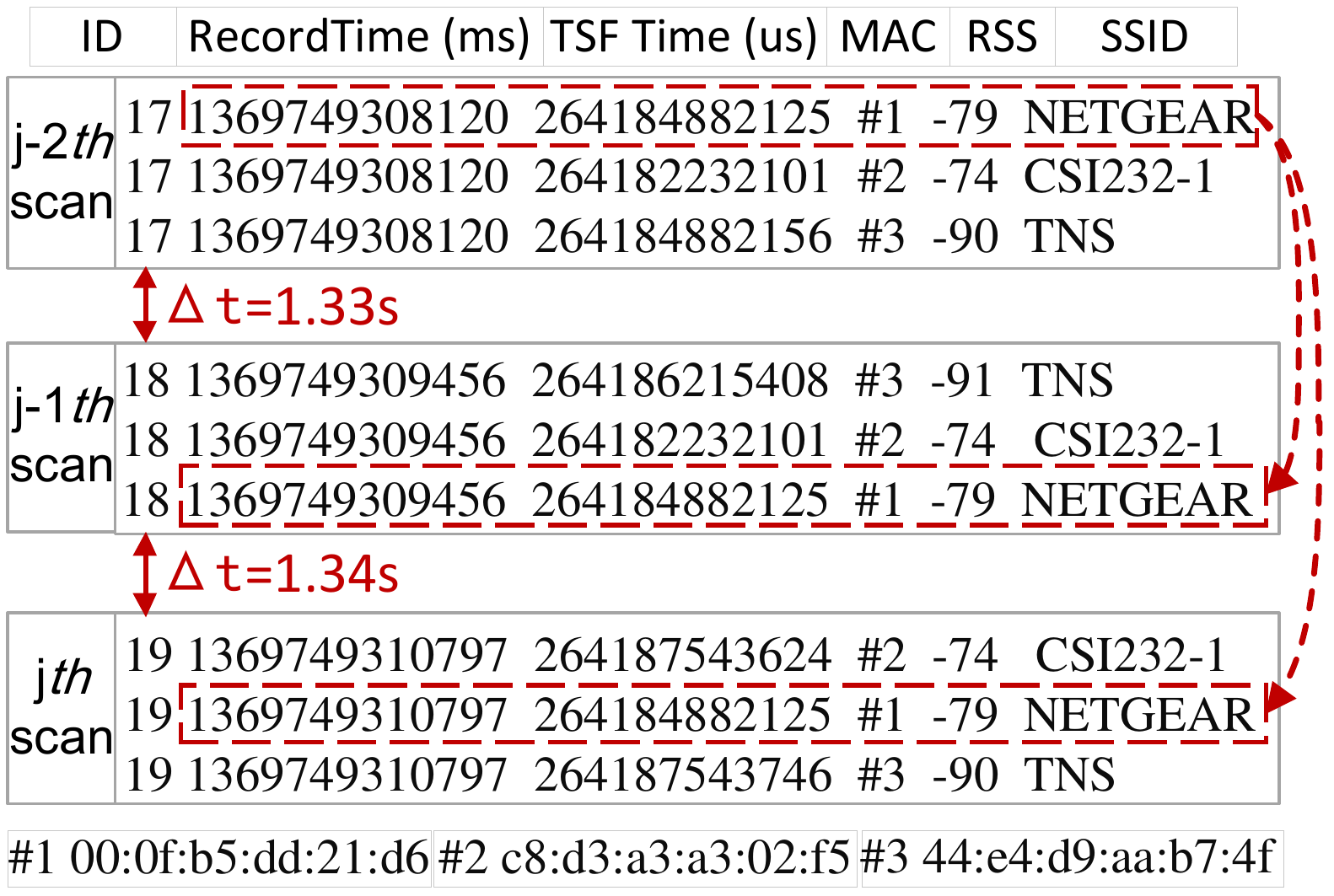}
  		\caption{Outdated fingerprints: A glance of scanning results from Android OS}
 		\label{fig:outdated_fingerprint}
	\end{minipage}	
\end{figure*}

The proliferation of mobile computing has spurred extensive interests in location-based services, leading to an urgent need for fine-grained location. The past decade has witnessed the conceptualization and development of various wireless indoor localization techniques, including WiFi \cite{bahl_radar_2000, youssef_horus_2005}, RFID \cite{ni_landmarc_2004, dina_rfid_2013}, acoustic signals \cite{tarzia_indoor_2011, liu2013guoguo}, ultrasound \cite{priyantha_cricket_2000,lazik_indoor_2012}, etc. 
Due to the wide deployment and availability of WiFi infrastructure, WiFi fingerprint-based indoor localization has become one of the most attractive localization techniques 
\cite{yang_locating_2012, shen2013walkie, wang_no_2012,liu_push_2012,rai_zee_2012, sen2013avoid}. 
Roughly speaking, a fingerprint-based scheme consists of two stages: site survey and fingerprint matching. During site survey (a.k.a calibration or war-driving), received signal strengths (RSS) from multiple WiFi access points (APs) are recorded at known locations to construct a fingerprint database. To locate a user, localization algorithms match his RSS measurements against the pre-labeled records and estimate his location to be the one with the best-fitted fingerprint.

\rev{There is generally a tradeoff between accuracy, ubiquity, and cost in designing a pervasive indoor localization system.} 
Accuracy has long been the primary challenge especially in mobile environments. 
\rev{Even schemes that have been reported to have very high accuracy in some instances}, e.g., \cite{youssef_horus_2005, lim_zero_2010, rai_zee_2012}, can experience rapid performance degradation in realistic environments, with median error consistently above 5 meters \cite{turner2011empirical}. 
\rev{In addition,} there are always unacceptably large tail errors, e.g., 10$\sim$20m or larger. Recent works \cite{turner2011empirical, liu_push_2012} found that 
large errors of prior works could range from 12 to around 40 meters. 
Mobility further deteriorates the performance especially for smartphone based methods. 
Efforts to gain high accuracy include to leverage physical layer information \cite{sen_you_2012} and  incorporate acoustic ranging \cite{liu2013guoguo, liu_push_2012}, among others.  These methods typically either rely on information unavailable on commodity smartphones, or resort to unrealistic cooperation among a dense crowd of peers. 
In this paper, we revisit the WiFi fingerprinting localization framework and ask whether it is possible to achieve accurate and robust fingerprint-based localization, especially for mobile phones, without degrading the ubiquity or increasing the costs. 

To investigate the root cause of limited localization accuracy, we conduct extensive experiments and uncover the following characteristics of WiFi fingerprint-based localization: 
1) APs have \rev{different discriminatory capabilities to fingerprint a specific location} since RSS changes are inversely proportional to the physical distance, subject to radio signal propagation laws. Intuitively, faraway APs may lead to large location estimation errors while close ones can help mitigate the location uncertainty. 
2) Biased RSS measurements caused by signal fluctuation and human body blockage may present themselves as outliers in fingerprint matching. Human body blockage to smartphones can remove line-of-sight and weaken the received signal by up to 10dB, thus greatly exaggerating the discrepancies of fingerprints measured from the same location. 
3) RSS measurements may be outdated due to hardware and software limitations of commodity wireless devices. In other words, latest reported RSS values could be duplicates of previous scans performed  several seconds ago. Considering user mobility, the outdated RSS could in fact be measurements done at a previous location, resulting in outdated fingerprints consisting of RSS measurements from multiple locations. In overlooking such outdated information, previous works directly compare the outdated fingerprints with those collected at a single location, incurring frequent fingerprint mismatches. 
The above are key reasons behind location errors of fingerprint-based schemes, especially in mobile environments; yet surprisingly they have not been adequately addressed in existing works.

With these observations in mind, we design \sysname~(named after \empha{D}iscrimination diversity, \empha{O}utdated \empha{R}SSs, and \empha{F}ingerprint \empha{in}consistency), a new fingerprint-based scheme for highly accurate localization. \sysname~includes three main components.
First, we quantitatively differentiate distinct AP's \rev{discriminatory ability w.r.t. a specific location}. \rev{APs with stronger ability are emphasized with more weights in fingerprint matching, while others are de-emphasized}. 
Second, noting fingerprint inconsistency, we apply a robust regression technique in fingerprint matching, in the hope of bounding the impact of RSS outlier values and ensuring accuracy under noisy measurements.
Finally, we propose phantom fingerprints that incorporate multiple \rev{fingerprints} in the fingerprint database to deal with the outdated RSS values. Phantom fingerprints are assembled according to the spatial constraints of outdated RSSs, which are derived by monitoring user mobility using smartphones' built-in inertial sensors.
Integrating these components, we design a uniform fingerprint similarity metric which further takes account of common AP ratio as a factor to mitigate \rev{errouneous matches of distant fingerprints}.

To validate our design, we implement \sysname~on commodity devices and conduct extensive experiments in multiple buildings. Experimental results demonstrate competitive performance of \sysname~to solutions based on physical layer information or on additional ranging techniques. In addition to the average accuracy of 2m, \sysname~significantly reduces large location errors by limiting the 95 percentile errors in 5.5m, both outperforming the state-of-the-art schemes like Horus by 56\% and 69\%, respectively. Using only the most essential RSS and requiring no extra hardware, we believe our approach takes an important step forward to accurate location estimation on smartphones with the prevalent WiFi infrastructure.

Our contributions are summarized as follows:
\begin{itemize}
	\item We uncover several crucial insights that explain the root cause of location errors but have not been adequately studied in existing literature. 
	\item We are the first to tackle the outdated RSS problem and the fingerprint inconsistency in WiFi fingerprint-based localization. We design an effective scheme for accurate and robust localization leveraging only the \rev{prevalent RSS}, requiring no extra information or additional hardware. 
	\item We implement a prototype system and conduct real world experiments in multiple buildings using commodity devices. \rev{In addition to the remarkable performance}, our method can be conveniently integrated in existing WiFi fingerprint-based localization systems.
\end{itemize}

The rest of the paper is organized as follows. Section \ref{sec:measurements} presents our preliminary measurements and basic observations. The method design is detailed in Section \ref{sec:design}, followed with the experiments and performance evaluation in Section \ref{sec:evaluation}. We discuss the state-of-the-art of indoor localization in Section \ref{sec:related-works} and conclude the paper in Section \ref{sec:conclusions}.

\begin{figure*}[t]
	\centering
		\subfloat[Outdated rates over different APs]{
			\includegraphics[width=\3figwidth]{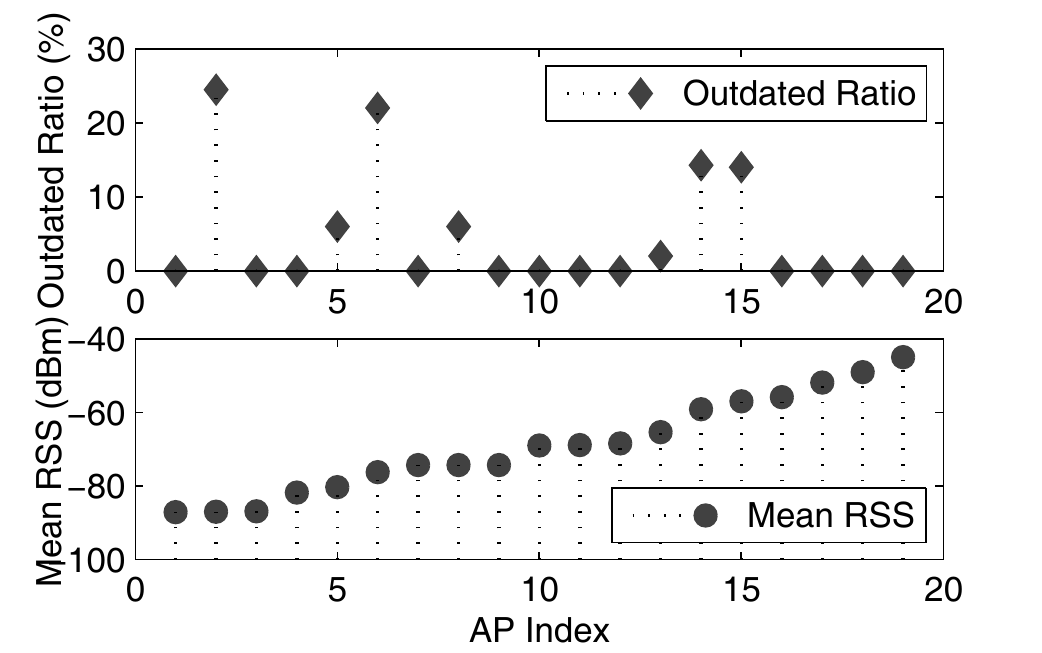}
			\label{fig:OutdatedRateOnAP}
		}
		\subfloat[Time delays of outdated RSSs]{
			\includegraphics[width=\3figwidth]{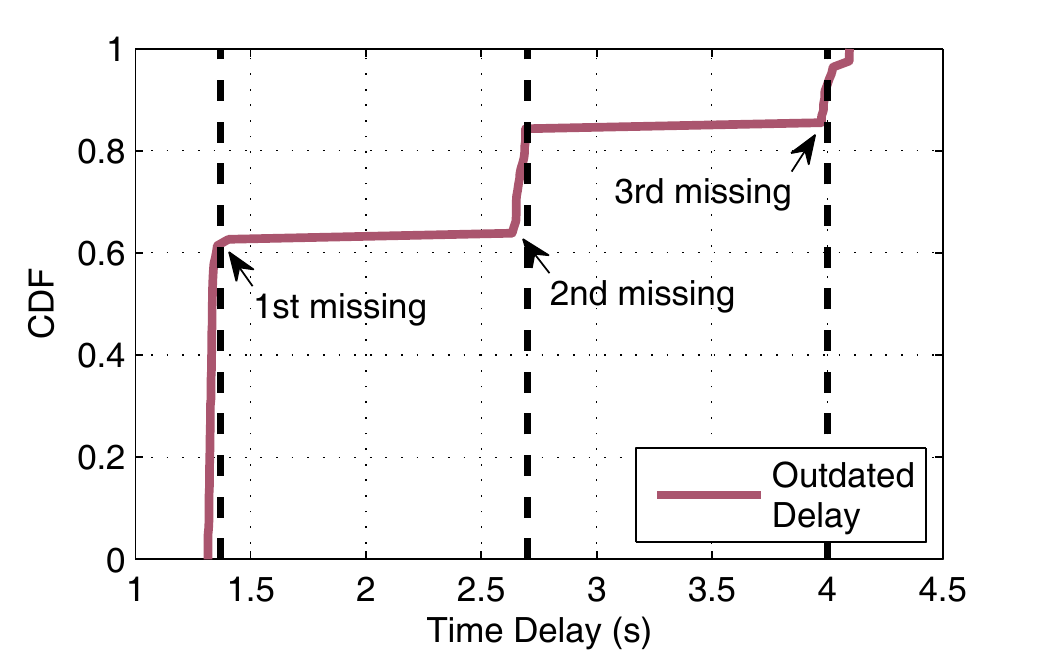}
			\label{fig:OutdatedDelayOnAP}
		}
		\subfloat[Time delays of outdated fingerprints]{
			\includegraphics[width=\3figwidth]{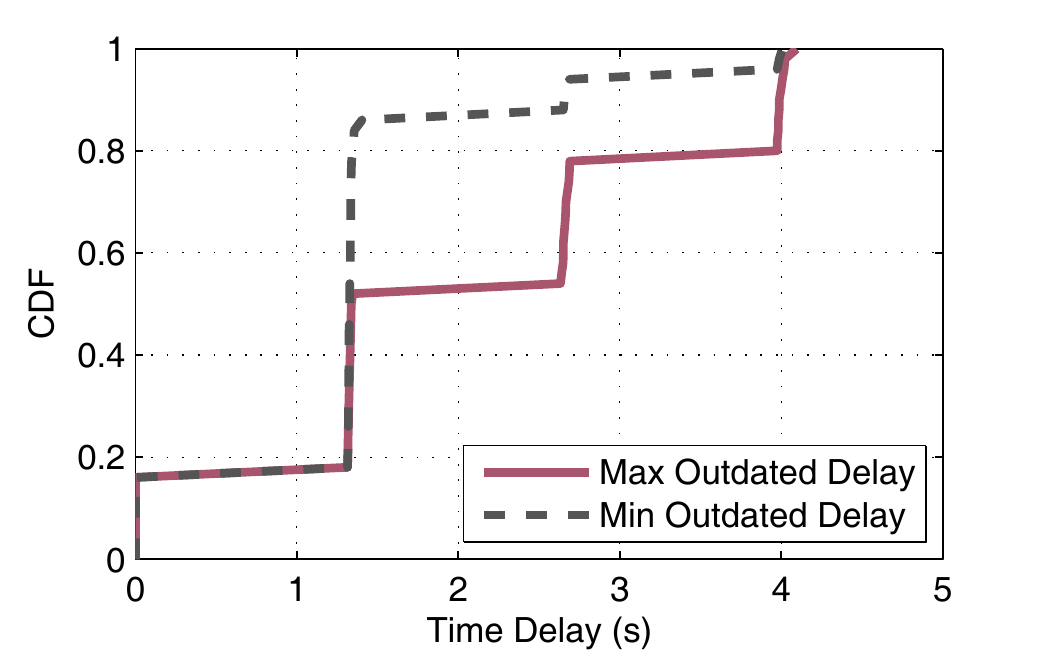}
			\label{fig:OutdatedDelayOnFin}
		}
		\hfill	
	\caption{Outdated RSS phenomenon: The reported RSSs in one fingerprint might be outdated.}
	\label{fig:OutdatedRSS}
\end{figure*}

\section{Preliminary and Measurements}
\label{sec:measurements}

In this section, we review the classical RSS fingerprinting problem and investigate fundamental characteristics of radio fingerprints through real measurements. Our preliminary results show some crucial features, which, having been largely overlooked in the past, shed light on how to achieve high accuracy of fingerprint-based localization.

\subsection{Problem Statement}
\label{sec:problem}

The working process of a typical fingerprint-based localization scheme consists of two stages: site survey and fingerprint matching. During site survey, wireless fingerprints (i.e., the set of RSS values from multiple APs) are measured and recorded at every location of interests. A fingerprint database (a.k.a radio map) is accordingly constructed, in which the fingerprint-location relationships are stored. To locate a user who sends a location query with his current RSS fingerprint, localization algorithms retrieve the fingerprint database and return the location of the matched fingerprint as the user's location estimation.

Denote a fingerprint as $\boldsymbol{f} = [f_i, i=1, \cdots, n]$, where $f_i$ is the RSS value of the AP $A_i \in \mathcal{A}$, the set of $n$ detectable APs appearing in $\boldsymbol{f}$. 
For two fingerprints $\boldsymbol{f}$ and $\boldsymbol{f'}$, denote the \term{RSS difference} (RSD) vector as $\bm{\delta} = [\delta_i, i=1, \cdots, p]$ where $\delta_i = |f_i-f_i'|$ indicates the RSD of AP $A_i \in \mathcal{A} \cup \mathcal{A}'$ in the two fingerprints and $p=|\mathcal{A} \cup \mathcal{A}'|$. Since $\boldsymbol{f}$ and $\boldsymbol{f'}$ do not necessarily contain identical sets of APs, we set $f_i$ ($f_i'$) to -100, the default minimum RSS value, if $A_i \notin \mathcal{A}$ ($\mathcal{A}'$). Let $\phi$ be the dissimilarity between  $\boldsymbol{f}$ and $\boldsymbol{f'}$, which, if measured by Euclidean distance, can be calculated as $\phi(\boldsymbol{f}, \boldsymbol{f'})=\lVert \boldsymbol{\delta} \rVert = \sqrt{\sum_{i=1}^p \delta_i^2}$. For all fingerprints stored in the fingerprint database $\mathcal{F}$, the goal of fingerprint matching is to find the fingerprint $\boldsymbol{f}^*$ that achieves the highest similarity with respect to the query fingerprint $\boldsymbol{f}$. Formally,
\begin{equation}
\boldsymbol{f}^* = \argmin_{\boldsymbol{f}_i \in \mathcal{F}} \phi(\boldsymbol{f}, \boldsymbol{f}_i).
\end{equation}
Then the user's location is estimated as the corresponding location $L(\boldsymbol{f}^*)$ of $\boldsymbol{f}^*$. Assumimg the true location of $\bm{f}$ is $L(\bm{f})$, the location estimation error is given by $\varepsilon = \lVert L(\bm{f})-L(\bm{f^*}) \rVert$.

\subsection{Observations}

\begin{obsv}[Discrimination Diversity]
APs have diverse discrimination capability to fingerprint a specific location, subject to inherent constraints of radio signal propagation. 
\end{obsv}
\term{Discrimination capability} is referred to as the ability of one AP to distinguish a specific location when using its RSS observations as fingerprints. Ideally, subject to the propagation law of wireless signals, RSS decays logarithmically with propagation distance $d$. More formally, $RSS \propto -\log (d)$, indicating that $\frac{\Delta RSS}{\Delta d} \propto -\frac{1}{d}$, where $\Delta RSS$ denotes the RSS change and $\Delta d$ is the corresponding distance change. \rev{In other words, an identical $\Delta RSS$ can imply a smaller distance change $\Delta d$ at closer locations, or a larger $\Delta d$ at faraway positions.} As shown in \figurename~\ref{fig:discrimination}, RSS variance of 1dB in value corresponds to vastly different changes in physical distance, depending on the specific $d$. Hence, faraway APs may cause large errors in location estimation, while close ones can conversely mitigate the errors. 
In a nutshell, distance changes indicated by RSS variances depend on the transmitter-receiver distance, leading to diverse discrimination capability across different locations.

%


\begin{obsv}[Fingerprint Inconsistency]
\label{obs:DI}
The majority of APs hold similar RSSs for fingerprints from the same/close locations while a small fraction may exhibit large differences due to environmental dynamics and human body blockages.
\end{obsv}
Location errors originate from unmatched fingerprints measured from the same/close locations. Our investigation on these fingerprints indicate that a majority of APs exhibit relatively stable RSSs even when these fingerprints are not matched. That is to say, the fingerprint dissimilarity (under certain metric such as Euclidean distance) is primarily produced by the drastically fluctuating RSSs of a small portion of APs, which is, however, obviously not caused by location changes, but probably stems from ambient dynamics and human body blocking effects \cite{welch2002effects,zhang_i_2011} especially in mobile environments. 


As shown in \figurename~\ref{fig:bodyblock}, signal strengths perceived by smartphones decrease significantly when the human body blocks the direct path of signal propagation, compared to when the user is facing the AP. These weakened RSS observations of blocked APs tend to deviate from the normal profiles, resulting in abnormal RSSs when compared with fingerprints measured during the training phase. 
Taking \figurename~\ref{fig:bodyblock} as an example, the normal RSS profile absent of body blockage is measured to be $\bm{f} = [-40, -65, -50]$. When a user is present and faces left, the right AP is blocked, resulting in a biased fingerprint $\bm{f}_{\mathrm{left}} = [-40, -65, -65]$ (for simplicity, we assume RSSs of unblocked APs remain unchanged). When facing right, the line of sight of the left AP is blocked and its RSS is correspondingly weakened, creating a fingerprint $\bm{f}_{\mathrm{right}} = [-52, -65, -50]$. Then comparing $\bm{f}_{\mathrm{left}}$ and $\bm{f}_{\mathrm{right}}$ with the normal $\bm{f}$ produces inconsistent RSD distributions $\bm{\delta}_{\mathrm{left}} = [0, 0, 15]$ and $\bm{\delta}_{\mathrm{right}} = [12, 0, 0]$, both generating abnormally larger fingerprint dissimilarity and ultimately leading to greater location uncertainty. Such inconsistency, however, has largely been overlooked in existing work but is ever-present especially for mobile users.


\begin{figure*}[t]
	\centering
	\begin{minipage}[t]{\3figwidth}\centering
		\includegraphics[width=1.0\textwidth]{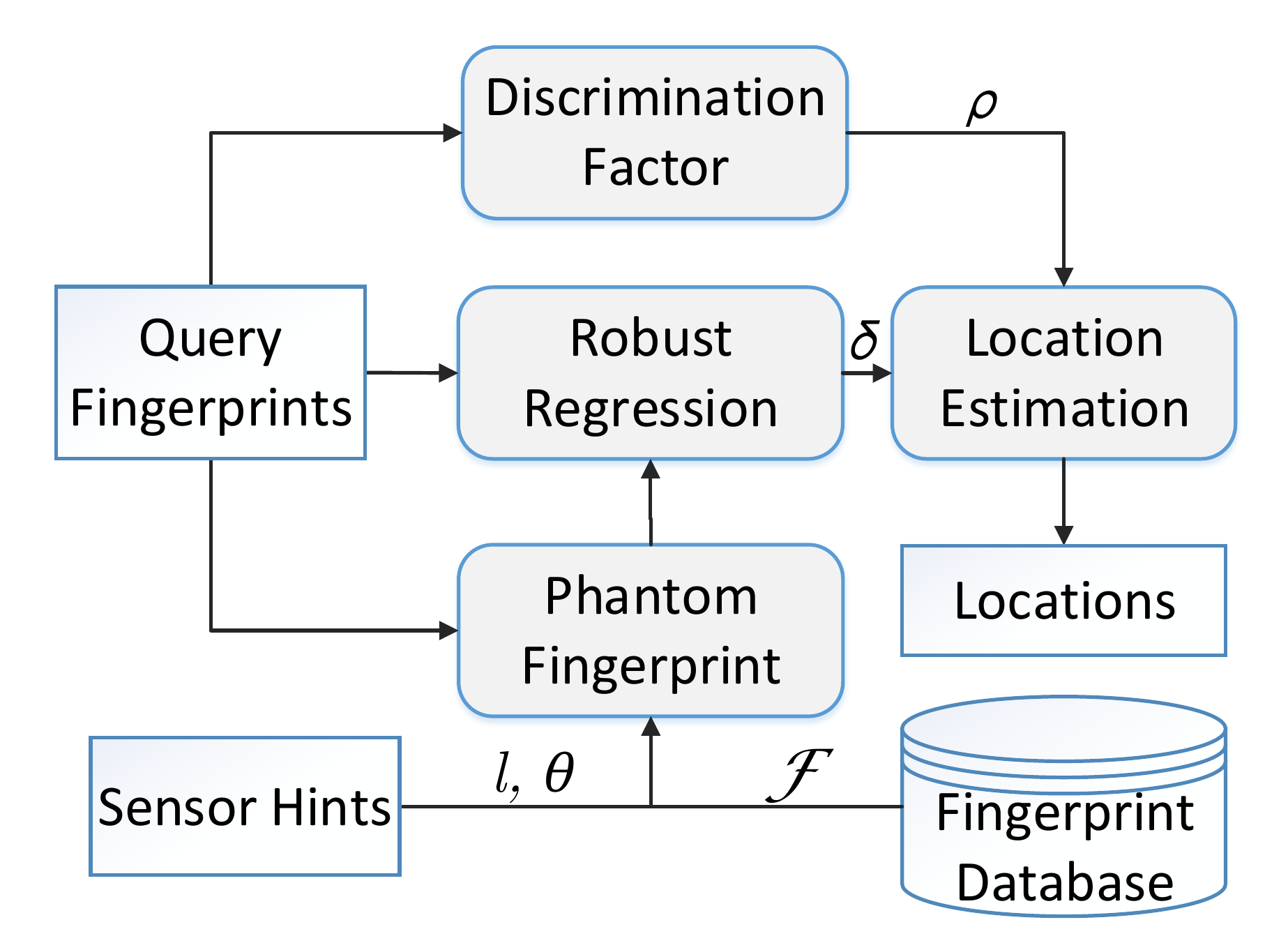}
  		\caption{System architecture}
  		\label{fig:architecture}
	\end{minipage}
	\hspace{0.2in}
	\begin{minipage}[t]{\3figwidth}\centering
		\includegraphics[width=\1figwidth]{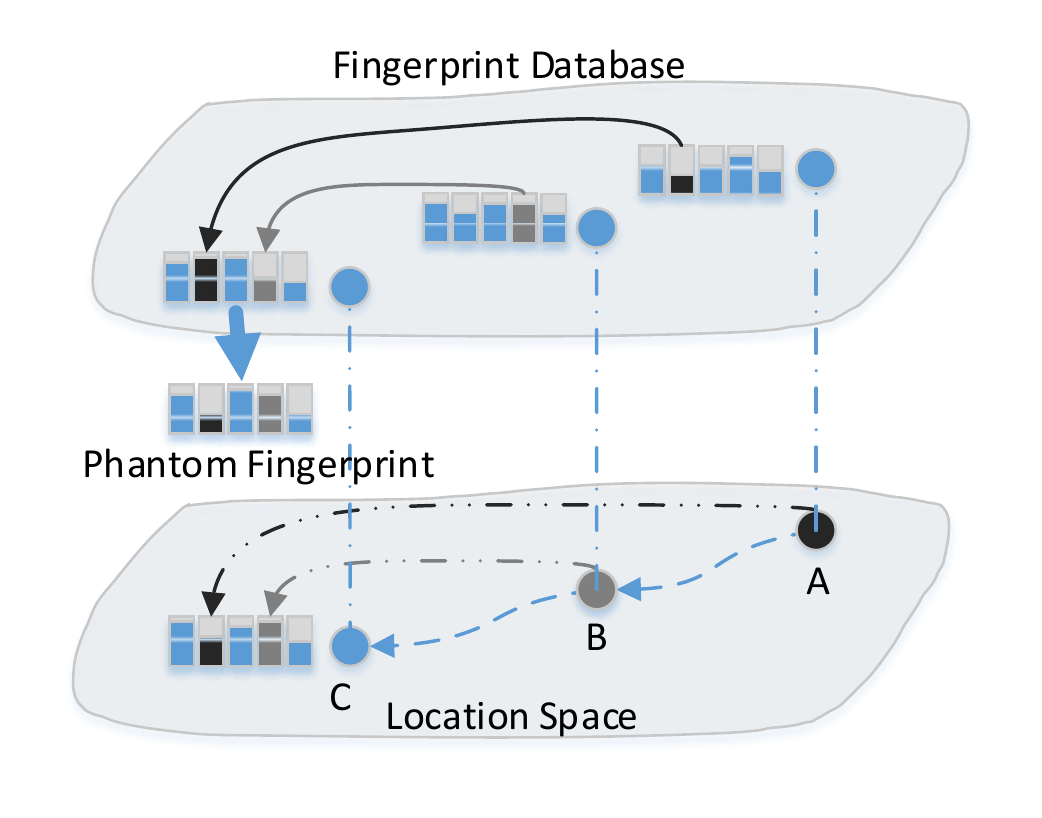}
  		\caption{Phantom fingerprint}
  		\label{fig:phantom-fingerprint}
	\end{minipage}
	\hspace{0.0in}
	\begin{minipage}[t]{\3figwidth}\centering
		\centering
  		\includegraphics[width=\1figwidth]{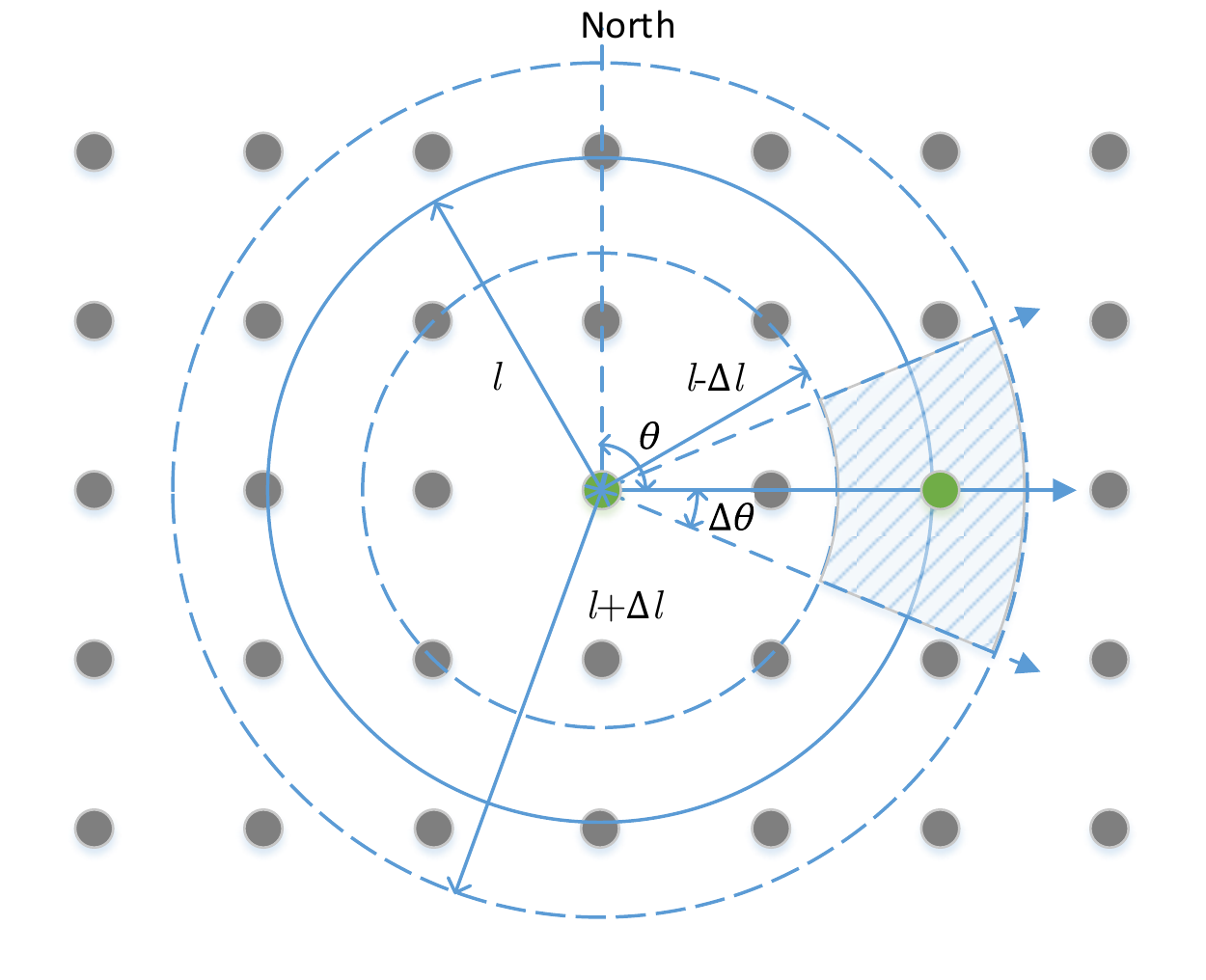}
  		\caption{Bequeathal locations}
 		\label{fig:phantom-location}
	\end{minipage}	\hspace{0.1in}
\end{figure*}

\begin{obsv}[Outdated Fingerprints]
The measured RSS might be outdated due to incomplete scanning results, caused by software and hardware restrictions.
\end{obsv}
Commodity smartphones acquire WLAN information in a \term{passive scanning} mode by listening to  periodic beacons from surrounding APs on all working channels. In this mode, the time a client stays on a channel is 100ms by default, which is specified by the 802.11 standard \cite{ieee2012wireless} and is equal to the default beacon interval. Consequently, the latency incurred in capturing the AP information for 2.4GHz WiFi is about 1,100ms since there are 11 available channels. In practice, it takes about 1$\sim$1.5 seconds for mainstream Android OS to complete a scan with commodity smartphones. Due to beacon conflicts and channel collisions, the beacon interval of 100ms cannot be always guaranteed, potentially resulting in some missed APs during a scan. However, to maintain quality of service, these missed APs can still appear in the scanning results by duplicating information from last several scans a few seconds ago.

As shown in \figurename~\ref{fig:OutdatedRSS}, a significant portion of APs experience high outdated rates, ranging from 2\% to 25\%. In particular, about 60\% of the outdated RSSs bear an outdated delay of 1.4s, while around 20\% and 15\% has a delay of 2.7s and 4s, respectively. \rev{Translated into fingerprints, \figurename~\ref{fig:OutdatedDelayOnFin} indicates that} 
over 80\% of fingerprints contain outdated RSSs, and for about 20\% the maximum delay time exceeds 4s. 

If a user is stationary, such outdated phenomenon has little impact on location fingerprinting since outdated RSSs are also measured from the same position. In mobile environments, however, users may have moved several meters away between consecutive scans, resulting in fingerprints comprised by RSS values that are actually observed at multiple locations, which we called \term{outdated fingerprints}. Previous works treat these outdated fingerprints as normal ones and compare them directly to those stored in the fingerprint DB, which are all collected at single locations. Obviously, matching fingerprints mixed from multiple locations to those from single positions may result in frequent fingerprint mismatches or even localization failures. 
In conclusion, serious outdated RSS measurements exist in WiFi-based location fingerprinting, and also contribute to location estimation error especially in a mobile environment.

In this study, we reconsider the RSS fingerprinting scheme based on these surprisingly overlooked observations. Specifically, we design \sysname, an accurate, robust, and practical indoor localization method which 1) quantitatively differentiates individual APs according to their location discrimination capability, 2) applies robust regression on inconsistent fingerprints, and 3) recombines phantom fingerprints to handle outdated RSS values in mobile environments. The following section details our design.


\section{Design Methodology}
\label{sec:design}

As illustrated in \figurename~\ref{fig:architecture}, the proposed solution includes a discriminatory policy, a phantom fingerprint assembling module, a robust regression procedure, and a normal fingerprint matching scheme.

\subsection{Discriminatory Policy}

Given that APs have diverse discrimination capability to fingerprint a specific location, it is inappropriate, and also unnecessary, to match two fingerprints with all APs equally involved. More accurate location estimations can be achieved by relying on the more discriminative APs, and limiting or even eliminating the influence of those fluctuating and distant ones. Toward this goal, we attempt to seek a discrimination metric that complies with physical constraints of signal propagation and simultaneously stays robust to RSS fluctuations.


To quantitatively differentiate each AP for a specific location, we define a \term{discrimination factor} by estimating the physical distance between the AP and the mobile client using the following Log-Distance Path Loss (LDPL) model \cite{rappaport1996wireless}:
\begin{equation}
P_d = P_{d_0} - 10\gamma\rev{\log}(\frac{d}{d_0}),
\end{equation}
where $P_{d_0}$ denotes the received power at a distance $d_0$ (which usually takes the value of one meter), $\gamma$ is the path loss exponent, and $P_d$ is the RSS in decibel measured at a distance of $d$ (in meters). 
Deriving the distance to AP $A_i$ from the LDPL model, we calculate its discrimination factor in fingerprint $\boldsymbol{f}_u$ to location $L_u$ as follows:
\begin{equation}
\rho_{i}^u = \frac{1}{d_{i}^u} = 10^{\frac{f_{u,i} - P_{d_0}}{10\gamma}},
\end{equation}
where $d_{i}^u$ is the estimated physical distance between $A_i$ and $L_u$. The rationale of using the reciprocal of physical distance lies in \rev{that it is consistent with the derivative of the LDPL equation, which indicates the RSS change $\Delta RSS \propto -\frac{1}{d}$. } 

While the exponential $\rho_{i}^u$ effectively discriminates different APs, it may also induce unnecessary matching errors in case of fluctuating RSSs. 
Consider one of the APs in a fingerprint that fluctuates to a very large value (e.g., -45dBm). In this case, the effects of other representative APs, which could hold considerable RSSs (e.g., up to -60dBm) and are thus discriminative, may become negligible since they can only get inappreciable factors three or four times smaller than the fluctuating AP. 
Hence to cope with noisy RSSs, we additionally incorporate a sigmoid function to retain the effects of most discriminative APs. Mathematically, $\rho_{i}^u$ is adjusted as follows:
\begin{equation}
\rho_{i}^u = \left\{
\begin{array}{ll}
	10^{\frac{f_{u,i} - P_{d_0}}{10\gamma}} & \textrm{if } f_{u,i} \leq f_0\\
	\frac{1}{a}\left( 1+e^{-2 \left(\frac{f_{u,i}+100}{10}-c \right)} \right)^{-1} & \textrm{otherwise}
\end{array}
\right.
\end{equation}
where all constant parameters $f_0$, a watershed RSS value, $a$ and $c$ can be determined by general empirical values and measurements. Afterwards, the discrimination factor is normalized such that $\sum_{k=1}^{n_u} \rho_{k}^u = 1$. 
The normalized $\rho_{i}^u$ then serves as a differential weight which will be attached to the RSD of $A_i$ between $\boldsymbol{f}_u$ and another fingerprint when computing their dissimilarity, as detailed in Section \ref{subsec:localization}. 


\subsection{Phantom Fingerprints}

As mentioned above, RSS of a specific AP reported in a query fingerprint may be outdated. If a user is stationary, this has little effect on fingerprint matching. 
In mobile environments, however, the outdated data may have been measured several seconds ago when the user was at another location. Subsequently, comparing fingerprint mixed with measurements from multiple locations with samples from one location can lead to false fingerprint matches. 


Intuitively, a query fingerprint consisting of RSS features of different locations should be matched with fingerprints recombined by measurements from multiple locations, which, however, are not directly available in the fingerprint database. In this sense, one needs to assemble special fingerprints, i.e., combinations of fingerprints from multiple locations, for matching, as shown in \figurename~\ref{fig:phantom-fingerprint}. 
These newly constructed fingerprints do not yet exist in the fingerprint database, and will be referred to as \term{phantom fingerprints}.

For a fingerprint $\bm{f} = [f_i, i=1, \cdots, n]$, denote the encountered timestamp of each AP $A_i$ in $\bm{f}$ by $t_i$. Recall \figurename~\ref{fig:outdated_fingerprint}, 
the scanning delay is typically longer than 1 second by our measurements while the differences of all APs' detected time in one fingerprint are usually small (indicated by the Time Synchronization Function\footnote{http://en.wikipedia.org/wiki/Timing\_Synchronization\_Function.} (TSF) timestamp). Hence, if the time difference between two APs in one fingerprint exceeds a certain length, e.g., 0.5s, then the earlier one is definitely outdated. In particular, for AP $A_k$, the outdated duration $\Delta t_k$ is computed by $\Delta t_k = \max\limits_{i=1,\cdots, n} t_i - t_k$. 

As illustrated in \figurename~\ref{fig:phantom-fingerprint}, assume that $f_k$ is actually the measurement of $A_k$ at a previous location, called \term{bequeathal location} (BL), where a user was present $\Delta t_k$ seconds ago. Further assume that the distance and direction from the BL to the user's current location is $\ell_k$ and $\theta_k$, respectively (we will describe how to compute $\ell_k$ and $\theta_k$ shortly). Then when comparing $\bm{f}$ with a candidate location, say, $L_z$, instead of directly computing the dissimilarity between $\bm{f}$ and $\bm{f}_z$, a sample fingerprint of $L_z$, we match it against the phantom fingerprints $\ef{\bm{f}}_z$ assembled from $\bm{f}_z$ and $\bm{f}_{BL(z)}$,  fingerprint from the BL. Concretely, the RSS value $f_k$ in $\bm{f}$ is replaced by that of the same AP in $\bm{f}_{BL(z)}$. In case of multiple outdated RSSs, all of them are replaced according to their individual BLs, finally resulting in a precise phantom fingerprint $\ef{\bm{f}}_z$.

The distance offset $\ell$ and direction $\theta$ can be estimated by dead reckoning method using smartphone built-in inertial sensors like accelerometer, gyroscope, and compass \cite{constandache_towards_2010, wang_no_2012, rai_zee_2012, shen2013walkie}. Specifically, we adopt the method proposed in \cite{wu2013footprints}, which counts steps as accurately as up to 98\%. The footsteps could then be converted to physical displacement by multiplying with the user's step length, which can be automatically tracked \cite{rai_zee_2012}. The direction, on the other hand, is estimated using gyroscope and compass as \cite{rai_zee_2012}. In the following, we demonstrate that although dead-reckoning may not be adequate for localization, it is sufficient for \rev{our purpose of estimating $\ell$ and $\theta$}. 

Due to noisy sensors and arbitrary human behavior, $\ell$ and $\theta$ cannot be 100\% accurately computed. To cope with the erroneous estimations, we introduce an error range for each of them, denoted as $\Delta\ell$ and $\Delta\theta$, respectively, and demonstrate that the procedure of choosing BLs can tolerate these errors gracefully. Mathematically, as shown in \figurename~\ref{fig:phantom-location}, potential BLs \rev{need to satisfy the condition} that their distances and directions to the candidate location are bounded in $[\ell-\Delta\ell, \ell+\Delta\ell]$ and $[\theta - \Delta \theta, \theta + \Delta\theta]$, respectively. The size of the shaded area is $S=\Delta\theta\big( (\ell+\Delta\ell)^2 - (\ell-\Delta\ell)^2)\big)=4\Delta\theta\ell\Delta\ell$. 
Assuming a location sample density of 2m$\times$2m and $\Delta\ell\leq 2$ meters, the minimal size $S_0$ to cover two sample locations should be at least $4\Delta\ell$ m$^2$. Thus, if $\Delta\ell\leq 2$ meters and $\Delta \theta < 1/\ell$, we have $S < S_0$, which means the shaded area covers at most one sample location, i.e., there is only one candidate BL. 
In practice, the maximal value of the missing delay $\Delta t$ is less than 5s (APs not seen for more than 5s would no longer be reported until being detected again next time). Thus, assuming a normal walking speed of 1.2 m/s, the distance offset can be at most 6 meters, resulting in a minimum value of $1/\ell$ of $\frac{1}{6}$. In other words, even though the distance and direction estimations are erroneous, we could identify a suspicious area and, with high probability, there is only one possible BL in the area, as long as the errors are in certain ranges ($\Delta\ell\leq 2$ meters and $\Delta\theta < \frac{1}{6}$). 
In case of multiple BLs (which is rare based on our measurements), the one closest to the center of the suspicious area (the shaded area shown in \figurename~\ref{fig:phantom-location}) is selected. Phantom fingerprints are then constructed by fingerprints from the candidate location and those from the BLs.

According to specific location sampling density, not all outdated RSS values need to be updated. Only RSSs with distance offsets $\ell$ exceeding half of the unit length of sampling grids should be replaced. If $\ell$ is less than half of the sampling distance (including being equal to 0 which means static user), fingerprints are merely treated in the traditional way.

\begin{figure*}[t]
	\centering	
		\subfloat[Office building.]{
			\includegraphics[height=1.5in]{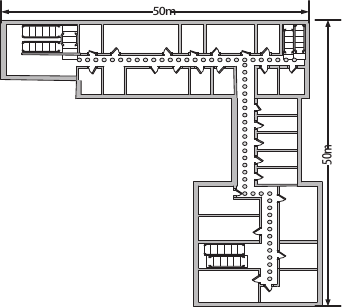}
			\label{fig:large_floorplan}
		}
		\hspace{0.5in}
		\subfloat[Classroom building.]{
			\includegraphics[height=1.5in]{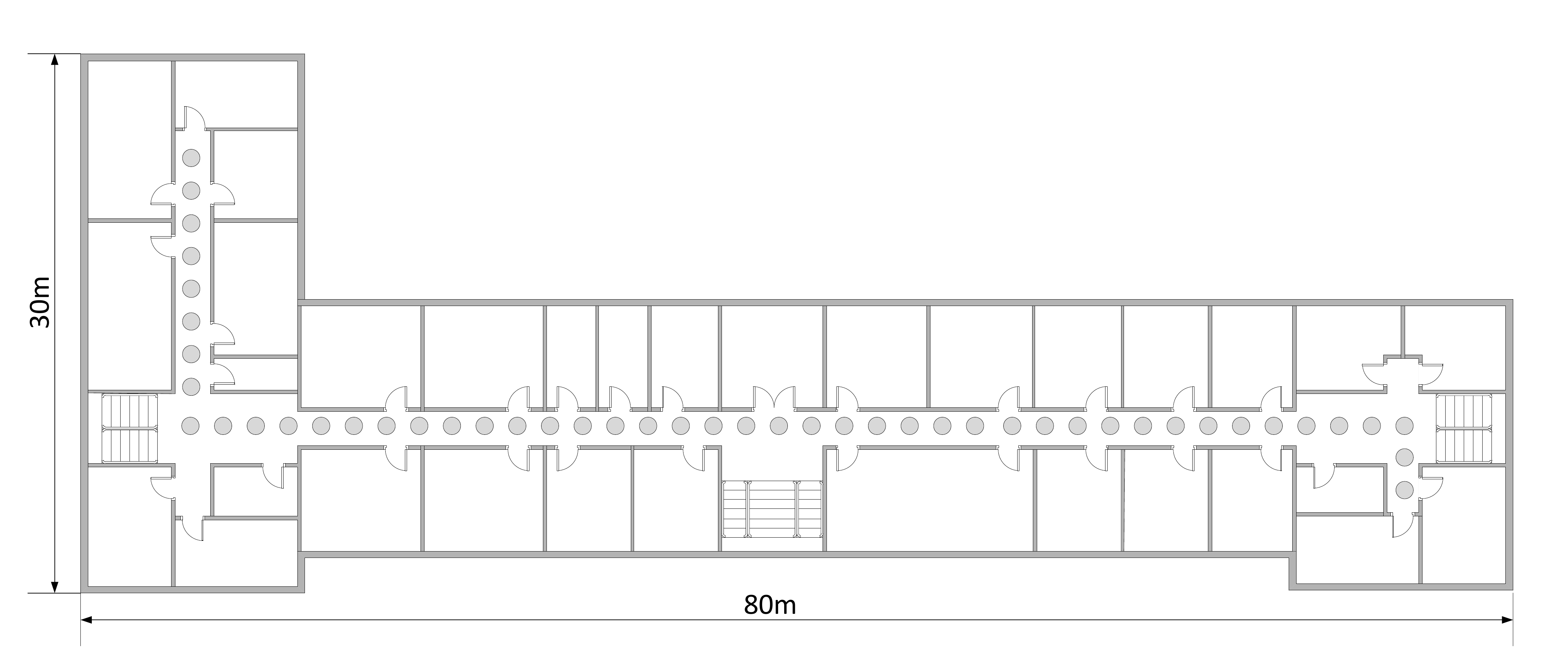}
			\label{fig:small_floorplan}
		}
		\hfill
	\caption{Experiment buildings.}
	\label{fig:floorplan}
\end{figure*}

\subsection{Robust Fingerprinting}

As we have observed, RSSs of one pair of fingerprints may contain outliers because of impaired measurements due to human body blockage. 
Since this is the primary cause of biased RSSs in mobile environments, only RSSs over a small portion of APs (that are blocked) may present outliers while most APs would remain consistent. Thus in this section, we propose to apply robust regression method on the inconsistent fingerprints, in the hope of bounding the influence of outlying measurements.



There are a large body of robust regression techniques, including $M$-estimator, $S$-estimator, $L$-estimator, etc \cite{rousseeuw2005robust}. Among them, we choose the most widely adopted Least Median of Squares (LMS) \cite{rousseeuw1984least} estimator due to its simplicity, effectiveness, and high breakdown point (0.5).

Given a query fingerprint $\bm{f}_s = [f_{s,i},1\leq i \leq p]$ and a sample fingerprint $\bm{f}_t = [f_{t,i}, 1\leq i \leq p]$, we adopt a simple linear regression model as follows:
\begin{equation}
y_{i} = \theta_1 x_i + \theta_2 + e_i, i=1,\cdots, p,
\end{equation}
where the response variables $\bm{y}$ are given by $\bm{f}_s$, while explanatory variables $\boldsymbol{x} = \bm{f}_t$. $\boldsymbol{e}=[e_1, \cdots, e_p]$ indicates the error term which is assumed to be normally distributed with zero mean and an unknown standard deviation $\sigma$. 
As the AP number $p$ is usually small, applying robust regression on insufficient observations does not always produce convincing statistical results. To obtain sufficient data for regression, we propose to compare the query fingerprint against all sample fingerprints corresponding to a candidate location, instead of a single averaged fingerprint. Specifically, for the candidate location $L=L(\bm{f}_t)$ with sample fingerprints $\mathcal{F}^L=\{\bm{f}_k^L, k=1,\cdots, m\}$, we simultaneously match $\bm{f}_s$ to all records in $\mathcal{F}^L$. 
In doing so, we acquire $mp$ observations, which can achieve the scale of hundreds since there are generally at least dozens of sample fingerprints for one location in the fingerprint database, and thus are \rev{sufficient} for statistical regression like LMS estimator. In this case, the explanatory variables $\bm{x}$ becomes $\bm{x} = [\bm{f}_1^L, \cdots, \bm{f}_m^L]_{1 \times mp}^T$ and correspondingly $\bm{y}$ is expanded as $\bm{y}=[\bm{f}_s, \cdots, \bm{f}_s]_{1\times mp}^T$. The regression model is thus rewritten as
\begin{equation}
y_{k,i} = \theta_1 x_{k,i} + \theta_2 + e_{k,i},
\end{equation}
where $i=1,\cdots, p, k=1,\cdots, m$, and $x_{k,i}$ and $y_{k,i}$ indicate the value of $f_i$ in $\bm{f}_k^L$ and $\bm{f}_s$, respectively. 
Applying LMS to the data $[\boldsymbol{x} \, \boldsymbol{y}]$ yields $\hat{\boldsymbol{\theta}}=[\hat{\theta}_1, \hat{\theta}_2]$ where the estimates $\hat{\theta}_i$ denote the regression coefficients. Multiplying $\boldsymbol{x}$ with these $\hat{\theta}_i$, we obtain the estimated values of $y_{i}$ as
\begin{equation}
\hat{y}_{k,i} = \hat{\theta}_1 x_{k,i} + \hat{\theta}_2.
\end{equation}
The LMS estimator is given by minimizing the median of squares of residuals as follows:
\begin{equation}
\min_{\hat{\boldsymbol{\theta}}} \mathop{\mathrm{med}}_{i,k} (y_{k,i}-\hat{y}_{k,i})^2.
\end{equation}


To determine whether a value $y_{k,i}$ is an outlier among all elements in $\boldsymbol{y}$, we compare the residual $r_{k,i} = y_{k,i}-\hat{y}_{k,i}$ to the \term{scale estimate} $\sigma^*$ defined by \cite{rousseeuw2005robust}. 
Then each $y_{k,i}$ is adjusted to $\tilde{y}_{k,i}$ as follows:
\begin{equation}
\tilde{y}_{k,i} = \left\{
\begin{array}{ll}
	y_{k,i} & \textrm{if } |r_{k,i}/\sigma^*| \leq 2.5\\
	\hat{y}_{k,i} & \textrm{otherwise}
\end{array}
\right.
\end{equation}
The boundary of 2.5 is an empirical value that has been \rev{suggested} by preliminary experience in the literature \cite{rousseeuw2005robust}. Accordingly, the RSS values of the query fingerprint $\bm{f}_s$ are regulated as 
$\tilde{f}_{s,i} = \frac{1}{m} \sum_k \tilde{y}_{k,i}$ and the RSDs $\boldsymbol{\delta}_{st}$ between $\bm{f}_s$ and $\bm{f}_t$ are thus tuned as 
$\tilde{\delta}_{st,i} = |\tilde{f}_{s,i} - f_{t,i}|$.

\subsection{Localization}
\label{subsec:localization}

Integrating all of the above components in a unified solution, we define a new metric as follows for uniform fingerprint dissimilarity judgment. 
\begin{equation}
h(\boldsymbol{f}_s, \boldsymbol{f}_t) = \Big(\sum_{i=1}^{p_{st}} (\rho_{i}^{st} \cdot \tilde{\delta}_{st,i})^2 \Big)^{\frac{1}{2}},
\end{equation}
where $p_{st}=|\mathcal{A}_s \cup \mathcal{A}_t|$ is the total number of distinctive APs in $\boldsymbol{f}_s$ and $\boldsymbol{f}_t$, and $\rho_{i}^{st} = \max\{\rho_{i}^{s}, \rho_{i}^{t}\}$ denotes the discrimination capability of AP $A_i$ for matching $\boldsymbol{f}_s$ and $\boldsymbol{f}_t$. 
Note that the RSD $\tilde{\delta}_{st,i}$ could also be given by other suitable metrics such as a probability estimation.
Realizing that fingerprints from closer locations share more common APs (or equivalently, fingerprints with very few common APs is unlikely to be from adjacent or same locations), the ultimate form of dissimilarity between two fingerprints $\boldsymbol{f}_s$ and $\boldsymbol{f}_t$ is expanded as follows:
\begin{equation}
\phi(\boldsymbol{f}_s, \boldsymbol{f}_t) = h(\boldsymbol{f}_s, \boldsymbol{f}_t) \cdot \frac{p_{st}}{q_{st}},
\end{equation}
where $q_{st}=|\mathcal{A}_s \cap \mathcal{A}_t|$ denotes the number of common APs in $\boldsymbol{f}_s$ and $\boldsymbol{f}_t$. 
With the above dissimilarity metric, the dissimilarity of two fingerprints with fewer common discriminative APs will be amplified. In case of no common APs ($q_{st} = 0)$, the dissimilarity will go to infinity, which eradicates the mismatch of two completely irrelevant fingerprints. 

\section{Experiments and Evaluation}
\label{sec:evaluation}

\begin{figure*}[t]
	\centering
	\subfloat[Accuracy in office building]{
		\includegraphics[width=\3figwidth]{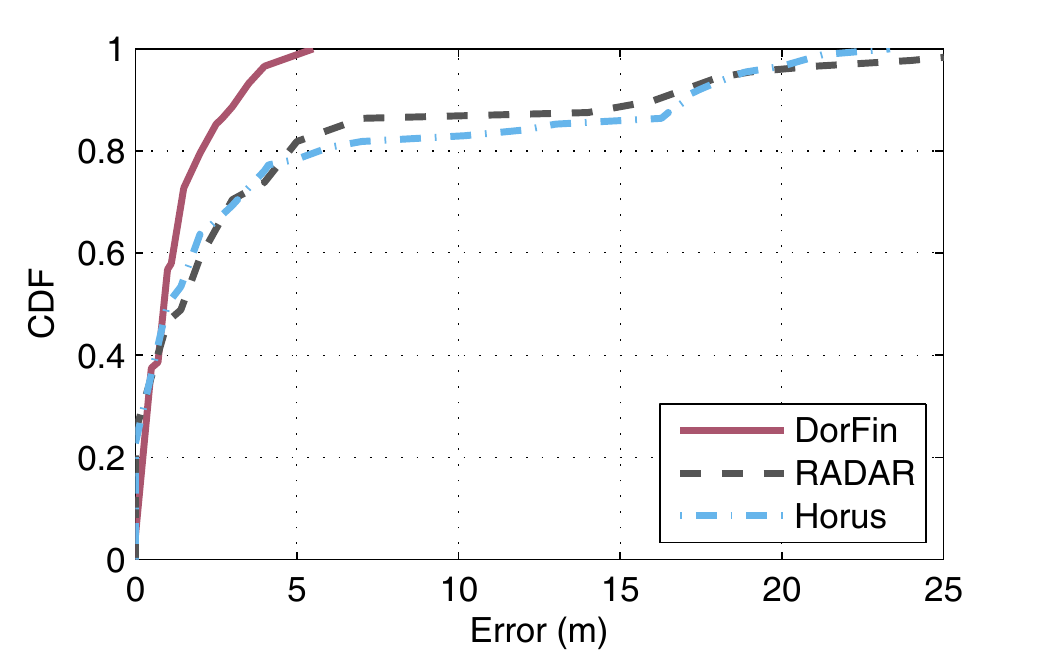}
  		\label{fig:accuracyDOR_10}
	}
	\hspace{0.01in}
	\subfloat[Accuracy in classroom building]{
		\includegraphics[width=\3figwidth]{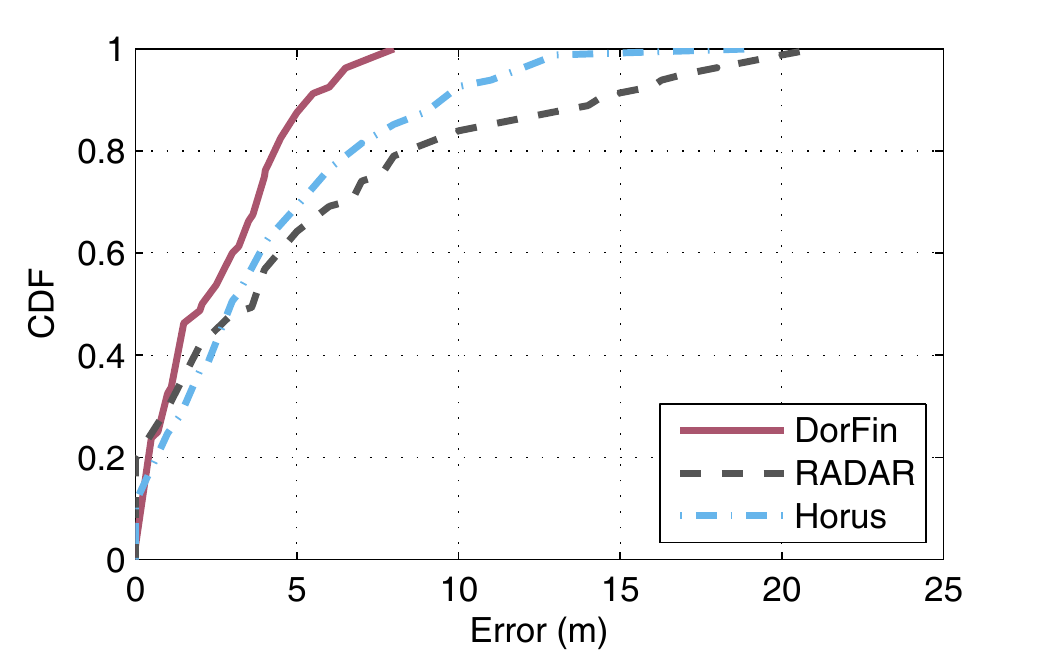}  		
  		\label{fig:accuracyDOR_10S}
	}
	\hspace{0.01in}
	\subfloat[Accuracy in mobile environments]{
  		\includegraphics[width=\3figwidth]{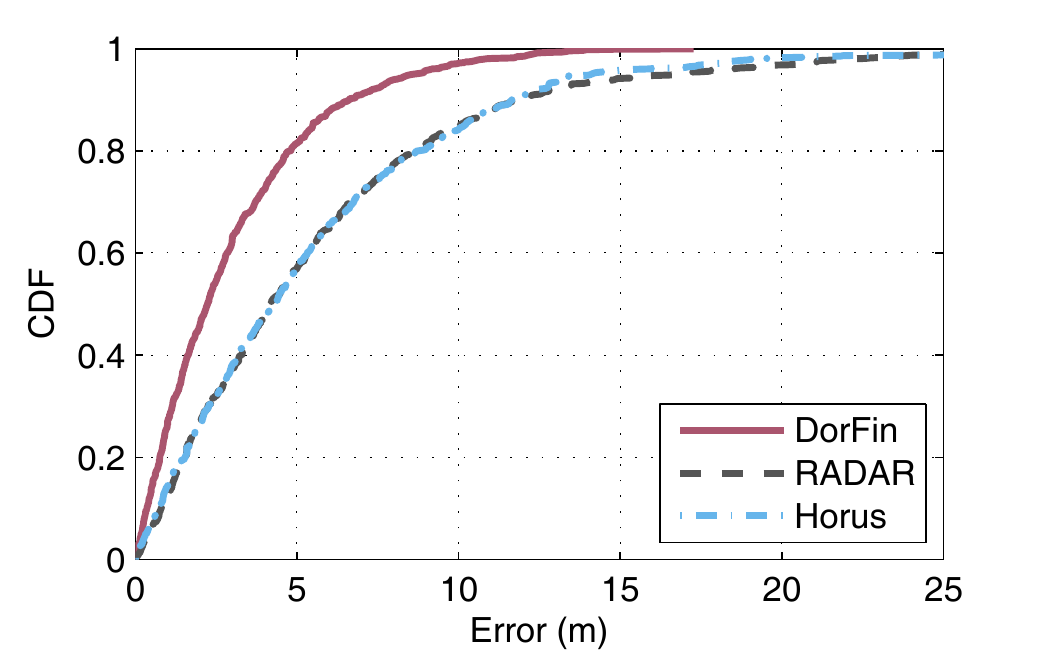}
 		\label{fig:accuracyDOR_PF}
	}
	\caption{Accuracy of \sysname}
	\label{fig:accuracyDOR}
\end{figure*}

\begin{figure*}[t]
	\centering
	\begin{minipage}[t]{\3figwidth}\centering
		\centering
  		\includegraphics[width=\fullfigwidth]{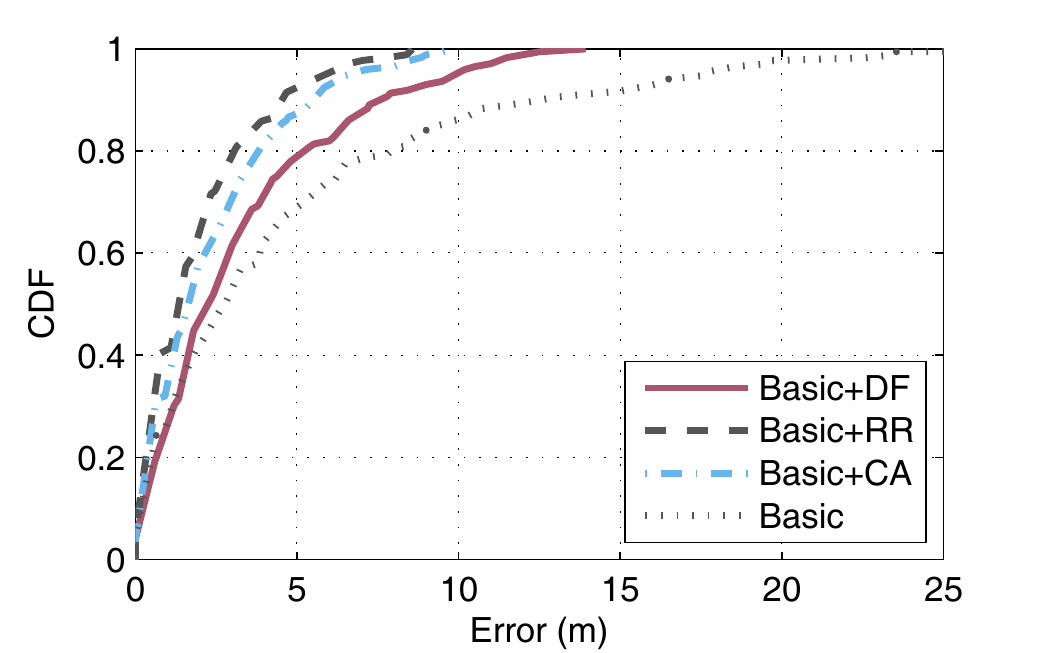}
  		\caption{Effects of individual modules}
 		\label{fig:accuracyDF_RR_CA}
	\end{minipage}	
	\hspace{0.01in}
	\begin{minipage}[t]{\3figwidth}\centering	
		\includegraphics[width=\fullfigwidth]{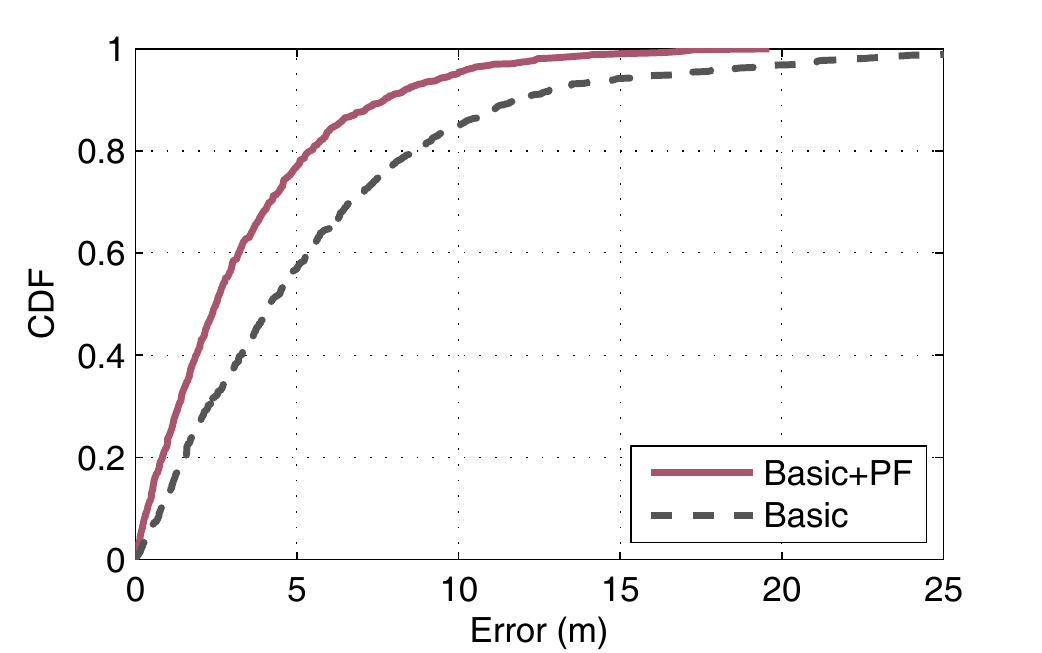}
  		\caption{Effect of phantom fingerprints}
  		\label{fig:accuracyPF}
	\end{minipage}
	\hspace{0.01in}
	\begin{minipage}[t]{\3figwidth}\centering	
		\includegraphics[width=\fullfigwidth]{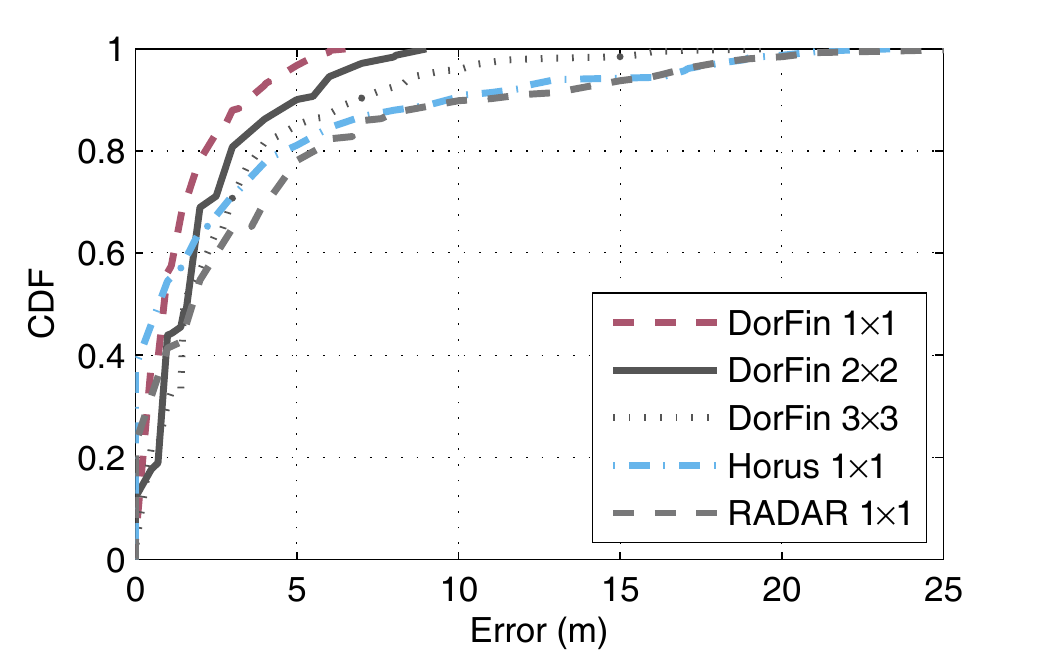}
  		\caption{Effect of sample density}
  		\label{fig:accuracyDOR_Density}
	\end{minipage}
\end{figure*}


\subsection{Experimental Methodology}
We prototype \sysname~on Google Nexus S and Nexus 4 phones which both run the mainstream Android OSs. We conduct the experiments in an office building and a classroom building on our campus, as shown in \figurename~\ref{fig:large_floorplan} and \figurename~\ref{fig:small_floorplan}, respectively. We manually sample areas of interests in both buildings and obtain a total of 83 sample locations in the classroom building and 90 locations in the office building. To construct the fingerprint database, we collect around 60 sample records at each location (which typically takes about 1 minute) by putting the phone on a portable desk. To obtain training data free of human body effects, no human is present around the desk when the mobile phone is collecting data. A high sampling density of 1m$\times$1m is used in site survey for extensive evaluation. Sparser data are then derived from these densely surveyed samples.

We consider both static and mobile cases for testing. For stationary cases, we collect query data by letting users record measurements at each location with their smartphones held in hand. For a mobile user, the smartphone measures wireless signals while the user is walking at a constant speed along a designated path with predefined start and end points. Note that the individual walking speed varies from user to user and from trace to trace. To obtain the ground truth locations of records along the moving trace, we compute user's walking speed by dividing the path length to the total time, and accordingly interpolate between the start and end points to obtain the location corresponding to each measurement based on their timestamps. In total, we collect static queries from around 200 locations, and gather over 20 mobile traces reported from different pathways, covering major areas of both buildings. 

A moving average filter is employed on the raw data to deal with noise. 
Location error between the ground truth and the estimated location (measured in Euclidean distance) is adopted for evaluation. In particular, we focus on the mean and 95th percentile localization errors.
To compare the performance to popular approaches in the literature, we employ a deterministic scheme RADAR \cite{bahl_radar_2000} and a probabilistic scheme Horus \cite{youssef_horus_2005}, two well-known methods of fingerprint-based localization. 
We choose RADAR and Horus for the purpose of confirming the performance improvements of \sysname~on pure RSS fingerprint-based schemes. To provide fair comparison, identical training and testing data are fed to Horus, RADAR, as well as our proposed approach.


\subsection{Performance Evaluation}

\subsubsection{Overall performance}
\figurename~\ref{fig:accuracyDOR} illustrates the localization error distributions seen by \sysname, Horus, and RADAR in different buildings and scenarios. \sysname~achieves mean accuracy of around 1.5m and 2.6m in office and classroom buildings respectively, consistently and substantially outperforming Horus and RADAR. Besides the promising average accuracy, \sysname~significantly reduces the maximum localization error. In both buildings, \sysname~bounds the 95th percentile errors to only about 3.7m and 6.7m respectively, while Horus and RADAR both generate location estimation errors larger than 15 meters under identical settings. Integrating all results in both buildings, \sysname~provides mean and 95th percentile errors of 2.0m and 5.5m, respectively. For comparison, the mean and 95th percentile errors of Horus are 4.4m and 17.9m while those of RADAR are 4.8m and 18m.

To examine the performance in mobile scenarios, we test the proposed approach on the mobile traces and report the integrated results.
Examining across all mobile traces, we observe that the average walking speed is 1.0m/s, while the maximum and minimum speeds are 1.5m/s and 0.6m/s respectively, all within regular range of human  walking speed.
As illustrated in \figurename~\ref{fig:accuracyDOR_PF}, despite slight drop in accuracy compared to the static cases, \sysname~maintains graceful performance in mobile cases, far superior to Horus and RADAR. Specifically, the average and 95th errors are about 3.0 meters and 8.5 meters, respectively. \rev{In comparison with RADAR and Horus, \sysname~decreases both errors by nearly 50\%.}  Even though the performance in mobile cases is not as good as static cases, the achieved accuracy remains comparable and promising. In addition, other complementary techniques such as path matching \cite{Yoon2013FM} can be integrated to further improve the accuracy for continuous localization.


\empha{Impact of sample density.} In the following, we further demonstrate that the suprior performance of \sysname~is attributed to the proposed approach instead of dense samples. 
As mentioned above, we sample the areas of interests with a density of 1m$\times$1m, which is relatively high for practical operations. To examine the performance with sparser sample locations, we perform \sysname~with training data of different sample densities (sample density is adjusted by sifting parts of the samples according to their locations). As shown in \figurename~\ref{fig:accuracyDOR_Density}, \sysname~preserves excellent accuracy even with sample densities of 2m$\times$2m and 3m$\times$3m. Specifically, with density of 2m$\times$2m, the mean and 95th percentile errors are still limited at 2.5m and 7.0m respectively, both better than those of Horus and RADAR with density of 1m$\times$1m.

In conclusion, \sysname~achieves remarkable performance in both stationary and mobile cases, with reasonable sample densities. 
To understand how each module of \sysname~contributes to the integral accuracy, we next perform an analysis across different modules.

\begin{figure*}[t]
	\centering
	\subfloat[Basic]{
		\includegraphics[width=\4figwidth]{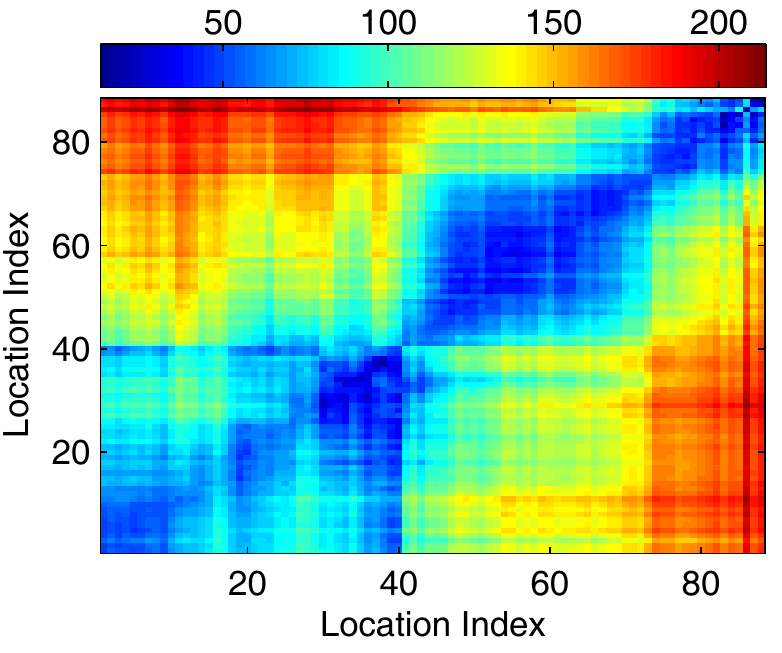}
  		\label{fig:confusionMatrix10Pure}
	}
	\subfloat[DF]{
		\includegraphics[width=\4figwidth]{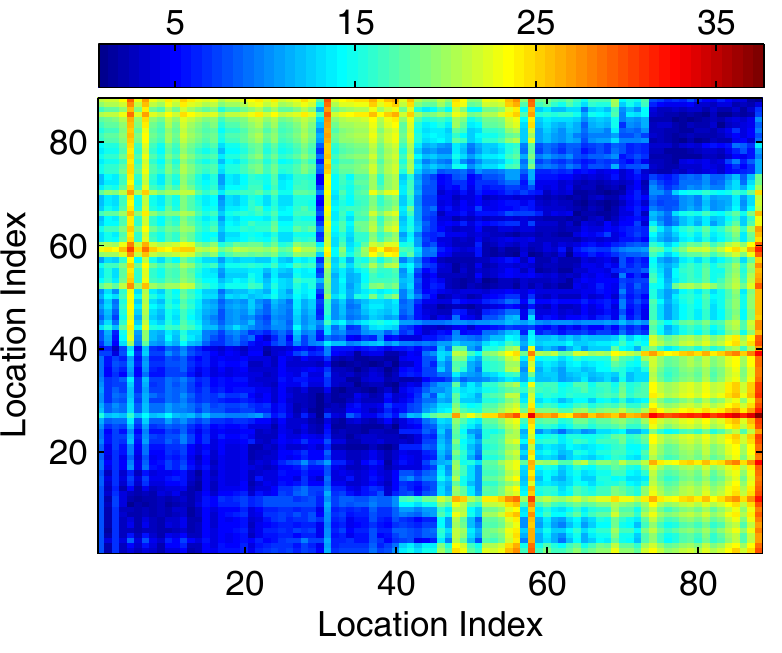}  		
  		\label{fig:confusionMatrix10DF}
	}
	\subfloat[CA]{
  		\includegraphics[width=\4figwidth]{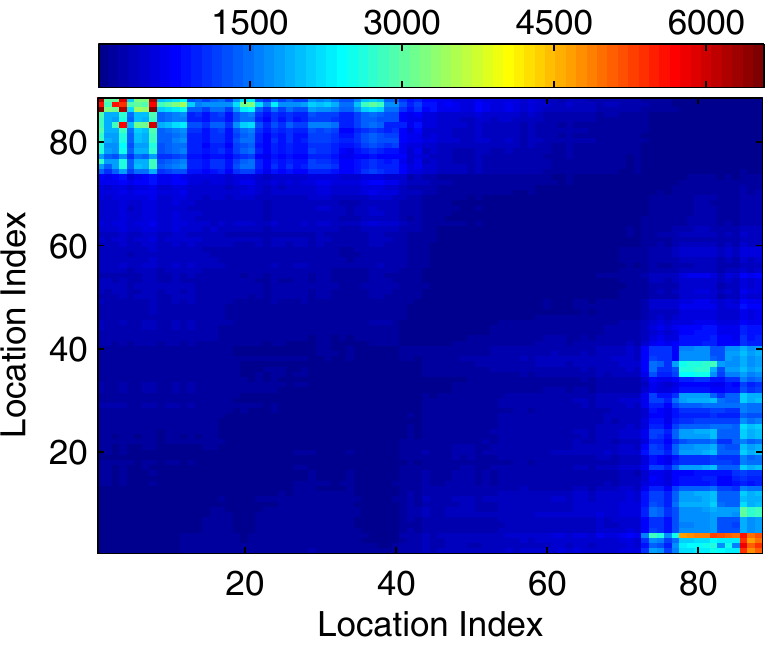}
 		\label{fig:confusionMatrix10CA}
	}
	\subfloat[RR]{
  		\includegraphics[width=\4figwidth]{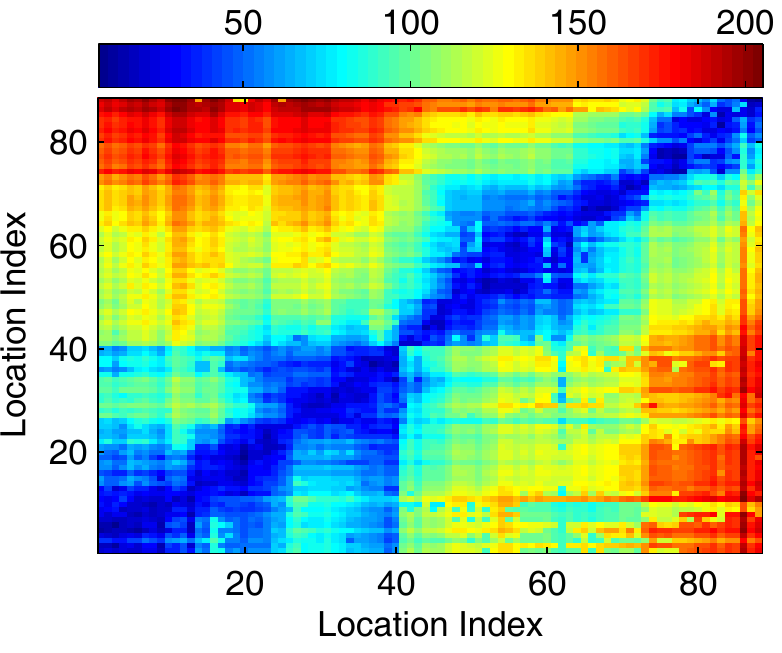}
 		\label{fig:confusionMatrix10RR}
	}
	\caption{Confusion matrix of similarity between query fingerprints and training fingerprints from 90 locations}
	\label{fig:confusionMatrix}
\end{figure*}

\subsubsection{Effect of Individual Modules}
To provide a clear analysis, we separately employ each module of \sysname, i.e., the discrimination factor (DF) module, the robust regression (RR) module, the common AP constraints (CA) module, and the phantom fingerprint (PF) module on the most basic nearest neighbor method (denoted as \term{Basic}) described in Section \ref{sec:problem} and evaluate the individual performance.

\empha{Effect of DF.}
We evaluate the impact of DF by using a set of empirical parameters to calculate the discrimination factor. Specifically, the path loss exponent is set to a typical value of 3 in indoor environments. The sigmoid function parameters $a$ and $c$ accordingly adopts the values of 4 and 4.3, respectively. As shown in \figurename~\ref{fig:accuracyDF_RR_CA}, DF limits the 95th percentile estimation error by about 40\%, while the average error is 1.5m lower than the Basic scheme, which has mean and 95th percentile errors of 5m and 17.5m. By placing more weight on more discriminatory APs and limiting those of the others, DF achieves the improvement by ensuring the similarity between fingerprints of close locations. This feature is also illustrated in \figurename~\ref{fig:confusionMatrix10DF}, which depicts the confusion matrix of fingerprint dissimilarity after employing the DF module (the confusion matrix of Basic scheme is shown in \figurename~\ref{fig:confusionMatrix10Pure}).  \rev{In addition, results from two buildings indicate that 
discrimination factors with uniform parameter settings can generate satisfactory results in different scenarios.} 

\empha{Effect of RR.} As shown in \figurename~\ref{fig:accuracyDF_RR_CA}, by employing RR over the Basic scheme, an average accuracy of 2.2m is achieved, with the corresponding 95th percentile accuracy of only 6m.
Evidently, the advantages of RR are the most significant among all modules by reducing the mean and 95th percentile errors by about 56\% and 65\% compared with the Basic scheme, respectively. 
\figurename~\ref{fig:confusionMatrix10RR} further depicts the corresponding dissimilarity matrix. 
Such results on RR confirm our observation that fingerprint inconsistency counts as a major cause of localization errors of fingerprint-based methods especially for smartphones.

\empha{Effect of CA.} \figurename~\ref{fig:accuracyDF_RR_CA} also demonstrates that the CA module is simple yet surprisingly effective. Incorporating the CA module with Basic scheme, the average and 95th percentile localization errors are reduced by about 50\% and 60\%, turning into 2.5m and 6.8, respectively. 
\figurename~\ref{fig:confusionMatrix10CA} shows the confusion matrix after weighing the fingerprint similarity by the common AP ratio. Dissimilarity of fingerprints from faraway locations is largely enlarged by the common AP ratio, while that of fingerprints from close locations is hardly affected (since close locations share more common APs).

\empha{Effect of PF.}
To examine the effectiveness of phantom fingerprints in dealing with outdated RSS measurements, we compare the performance of the Basic method on mobile data with and without constructing phantom fingerprints.
As depicted in \figurename~\ref{fig:accuracyPF}, the average and 95th percentile errors decrease from 3.9m and 10.4m to 2.4m and 6.9m respectively when the sample fingerprints are appropriately replaced with phantom fingerprints.
With these results, it is of interests to examine to what extent the measured RSSs and further the entire fingerprints are outdated. 
As we observed, over 11\% of RSS measurements are outdated in our experiment data. Furthermore, almost every fingerprint undergoes outdated RSSs. 
In particular, there always exist large outdated distances ranging from 2m to 6m in most fingerprints. 
The improvements gained by using PF validate quite convincingly our observation that the outdated RSS values can lead to location estimation errors in mobile environments.

Building on these components, \sysname~produces promising accuracy levels even in mobile environments that are competitive with that achieved by leveraging physical layer information \cite{sen_you_2012, sen2013avoid} or introducing extra ranging techniques \cite{nandakumar_centaur_2012, liu_push_2012} (both with mean accuracy of about 1m$\sim$3m). Requiring no hardware modification, \sysname~can enhance existing WiFi positioning systems.


%


\section{Related Works}
\label{sec:related-works}

In the literature of indoor localization, many techniques have been proposed in the past two decades. The state-of-the-art generally falls into two categories: fingerprint-based and ranging-based.

\textbf{Fingerprint-based techniques.} A large body of indoor localization approaches adopts fingerprint matching as the basic scheme for location estimation. 
Researchers have explored diverse signatures including WiFi \cite{youssef_horus_2005}, RFID \cite{ni_landmarc_2004}, FM radio \cite{Yoon2013FM}, acoustic \cite{tarzia_indoor_2011}, magnetism \cite{chung_indoor_2011}, etc. 
Among various signatures used, WiFi based scheme has been the most attractive. 

Smartphones with various built-in sensors have been leveraged in fingerprint-based localization to reduce or eliminate site survey efforts. Examples include LiFS \cite{yang_locating_2012}, unloc \cite{wang_no_2012}, Zee \cite{rai_zee_2012}, Walkie-Markie \cite{shen2013walkie}, etc. 
To provide better accuracy, sophisticated probability models and advanced machine learning techniques have been employed \cite{youssef_wlan_2003, youssef_handling_2004}. 
The study \cite{turner2011empirical} validates a broad range of approaches in a realistic environments and reports that median errors of prior work are consistently greater
than 5 meters and, counter-intuitively, that simpler algorithms frequently outperform more sophisticated ones. 
Realizing that large errors always exist due to possibly faraway locations with similar WiFi signatures, authors in \cite{nandakumar_centaur_2012, liu_push_2012} attempt to incorporate acoustic ranging in WiFi fingerprinting to limit the large tail errors. 
Although significant improvements are achieved, these approaches either rely on ranging among a dense crowd of users or require calibrating additional information. 
To completely bypass the instability of RSS, physical layer information, e.g., Channel State Information (CSI), is introduced and achieves an accuracy of $\sim$1m \cite{sen_you_2012}, but at the cost of ubiquity degradation (since CSI is unavailable on most commodity smartphones). 
To reduce computational complexity, different criteria for AP's discriminatory ability such as InfoGain \cite{chen2006power} and MaxMean \cite{youssef_wlan_2003} have been proposed to choose a subset of APs. However, they are only used for AP selection, instead of attaching to each AP for fingerprint matching.

\textbf{Ranging-based techniques.} These schemes calculate locations based on geometrical models rather than search for best-fitted signatures from pre-labeled reference database. The prevalent LDPL model, for instance, builds up a semi-statistical function between RSS values and RF propagation distances \cite{lim_zero_2010,chintalapudi_indoor_2010}. These approaches trade measurement efforts for the cost of decreasing localization accuracy. EZ \cite{chintalapudi_indoor_2010} employs a modeling method assuming no knowledge of physical layout or AP locations, and reports median error of 7 meters. 
Apart from RSS-based ranging, CSI is recently used to obtain for highly accurate distance and angle estimation \cite{sen2013avoid}. 
Acoustic ranging is also employed for fine-grained indoor localization, such as Centour \cite{nandakumar_centaur_2012}, Guoguo \cite{liu2013guoguo}, etc. 

Different from previous works that introduce additional information or extra signal sources for high accuracy, we identify the root causes of location errors in WiFi fingerprint-based localization for mobile devices, which have been largely overlooked in the literature. In addition, the proposed scheme achieves high accuracy with merely the prevalent RSS, thus is more amenable for practical applications.

\section{Conclusions}
\label{sec:conclusions}

While WiFi fingerprint-based localization acts as the dominant scheme in indoor localization, the accuracy challenge remains a primary concern. In this paper, we identify several crucial causes of localization errors in fingerprint-based schemes. 
These observations then lead us to the design of a new WiFi fingerprinting scheme which successfully reduces the mean and 95th percentile location errors to 2 meters and 5.5 meters, without degrading  ubiquity nor increasing the costs. Our approach marks a significant progress in RSS fingerprint-based indoor localization, especially for smartphones, and sheds lights on practical deployment in the real world.

\balance

\bibliographystyle{IEEEtran}
\bibliography{./bib/CS-Woo}

\end{document}